\documentclass{emulateapj}

\usepackage{float}
\usepackage{amsmath}
\usepackage{epsfig,floatflt}
\usepackage{subfigure}

\begin{document}

\title{Velocity dispersions in a cluster of stars: How fast could
  Usain Bolt have run?}

\author{H. K. Eriksen\altaffilmark{1,5,6}, J. R. Kristiansen\altaffilmark{2,5},
  \O. Langangen\altaffilmark{3,5} and I. K. Wehus\altaffilmark{4,7}}

\altaffiltext{1}{email: h.k.k.eriksen@astro.uio.no}
\altaffiltext{2}{email: j.r.kristiansen@astro.uio.no}
\altaffiltext{3}{email: oystein.langangen@astro.uio.no}
\altaffiltext{4}{email: i.k.wehus@fys.uio.no}

\altaffiltext{5}{Institute of Theoretical Astrophysics, University of
Oslo, P.O.\ Box 1029 Blindern, N-0315 Oslo, Norway}

\altaffiltext{6}{Centre of
Mathematics for Applications, University of Oslo, P.O.\ Box 1053
Blindern, N-0316 Oslo}

\altaffiltext{7}{Department of Physics, University of Oslo, P.O.\ Box
1048 Blindern, N-0316 Oslo, Norway}

\date{Received - / Accepted -}

\begin{abstract}
  Since that very memorable day at the Beijing 2008 Olympics, a big
  question on every sports commentator's mind has been ``What would
  the 100 meter dash world record have been, had Usain Bolt not
  celebrated at the end of his race?'' Glen Mills, Bolt's coach
  suggested at a recent press conference that the time could have been
  9.52 seconds or better. We revisit this question by measuring
  Bolt's position as a function of time using footage of the run, and
  then extrapolate into the last two seconds based on two different
  assumptions. First, we conservatively assume that Bolt could have
  maintained Richard Thompson's, the runner-up, acceleration during
  the end of the race. Second, based on the race development prior to
  the celebration, we assume that he could also have kept an
  acceleration of 0.5 m/s$^2$ higher than Thompson. In these two
  cases, we find that the new world record would have been
  $9.61\pm0.04$ and $9.55\pm0.04$ seconds, respectively, where the
  uncertainties denote 95\% statistical errors. 
\end{abstract}

\keywords{popular science --- image analysis --- Beijing 2008}

\maketitle

\section{Introduction}

On Saturday, August 16th 2008, Usain Bolt shattered the world record
of 100 meter dash in the Bird's Nest at the Beijing Olympics 2008. In
a spectacular run dubbed ``the greatest 100 meter performance in the
history of the event'' by Michael Johnson, Bolt finished at 9.69
seconds, improving his own previous world record from earlier this
year by 0.03 seconds. However, the most impressive fact about this run
was the way in which he did it: After accelerating away from the rest
of the field, he looked to his sides when two seconds and 20 meters
remained, and when that noting he was completely alone, he started
celebrating! He extended his arms, and appeared to almost dance along
the track.

\begin{figure*}
\begin{center}
\mbox{\epsfig{file=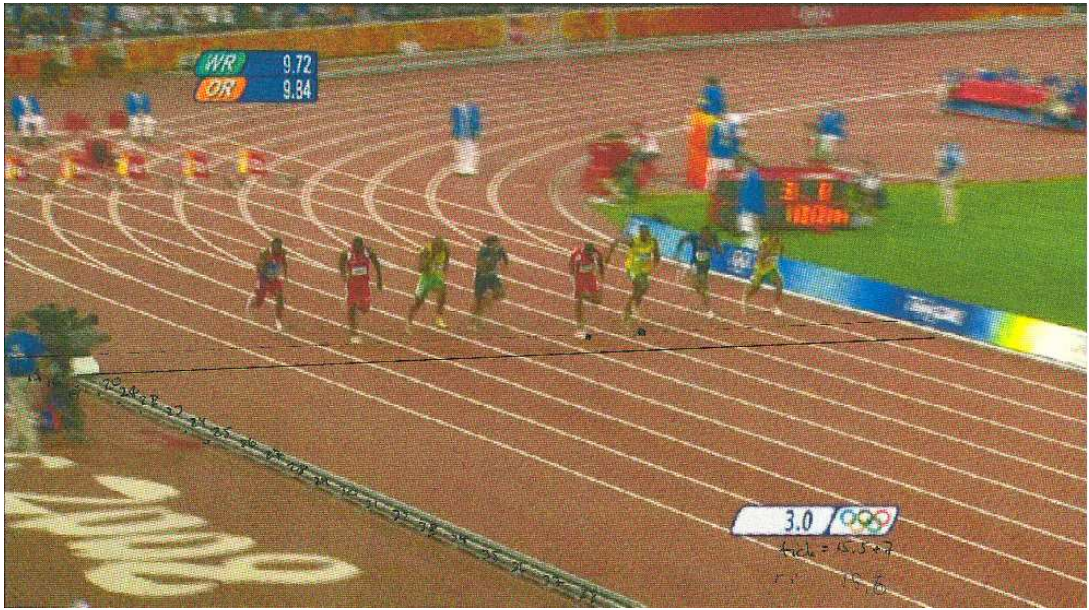,width=0.8\linewidth,clip=}}
\end{center}
\caption{Example screen shot used to estimate the runners' position as
  a function of time.}
\label{fig:screenshot}
\end{figure*}

Despite this, he broke the world record by 0.03 seconds. But, needless
to say, this celebration left spectators and commentators all over the
world wondering about one big question: What would the world record
have been if he had \emph{not} celebrated the last 20 meters? Bolt's
coach, Glen Mills, recently suggested at a press conference of the
Golden League tournament in Z{\"u}rich, that the record could have
been 9.52 seconds, or even better.

We wanted to check this for ourselves, by attempting to measure Bolt's
position as a function of time, and extrapolate from the dynamics
before the celebration began, into the last two seconds of the race.
Based on (hopefully) reasonable assumptions, we could then obtain an
estimate of the new world record.

In this paper we analyze footage of the run obtained from various web
sites and the Norwegian Broadcasting Corporation (NRK), with the goal
of estimating this ``hypothetical'' world record. The main technical
difficulty in performing this analysis lies in obtaining accurate
distance measurements as a function of time for each
runner. Fortunately, this task is made considerably easier by the
presence of a moving camera mounted to a rail along the track. This
rail is bolted to the ground at regular intervals, and thereby
provides the required standard ruler. Using the methods detailed in
the following sections, and properly taking into account all major
sources of statistical uncertainty, we believe that our measurements
are sufficiently accurate and robust to support interesting
conclusions. 

\section{Method}
\label{sec:method}

Our analysis is based on the following simple steps:
\begin{enumerate}
\item We first obtained several different videos of the race from the
  Internet (NBC and BBC) and the Norwegian Broadcasting Company (NRK),
  and printed out $\sim30$ screen shots at different times from these.
\item We then constructed a standard ruler by counting the total
  number of bolts (called ``ticks'' in the following) on the rail of
  the moving camera along the 100 meter track (see Figure
  \ref{fig:screenshot}). We assumed the distance between these to be
  constant.
\item Next, we drew lines orthogonal to the track, using whatever
  means most accurate for a given screen shot. For early and late
  frames, lines in the actual track itself (e.g., starting and
  finishing lines) were most useful, while for intermediate frames,
  the lower right edge of the camera mount was utilized (Figure
  \ref{fig:screenshot}).
\item For a given frame, we then read off the positions of Usain Bolt
  and Richard Thompson, the runner-up, with the ruler, and recorded
  these together with the time from the screen clock.
\item Next, we assigned an uncertainty to each distance measurement,
  by estimating how many ticks we believed we were off in a given
  frame. For later frames, when the camera angle is almost orthogonal
  to the track, this uncertainty is smaller than in the beginning of
  the race because of the camera perspective.
\item Based on these uncertainties, we fitted a smooth spline with
  inverse variance weights to the data. This provided us with a smooth
  approximation to the runners' positions as a function of time, and
  also with the first and second derivatives, i.e., their speeds and
  accelerations.
\item To make the projections, we consider two cases: First, we
  conservatively assume that Bolt would have been able to keep up with
  Thompson's acceleration profile in the end race after 8 seconds of
  elapsed time, and project a new finishing time. Second, given his
  clearly stronger acceleration around 6 seconds, we also consider the
  case in which he is able to maintain a $\Delta a =$0.5
  m/$\textrm{s}^2$ higher acceleration than Thompson through to the
  end.
\end{enumerate}

The final goal is the new projected world record, which is found by
extrapolating the resulting motion profile to 100 meters. We also
estimate the uncertainty in this number by repeating the above
analysis 10\,000 times, each time adding a random fluctuation with
specified uncertainties to each time and tick count. 

\begin{deluxetable*}{ccccccl}
\tablewidth{0pt}
\tablecaption{Position as a function of time for Bolt and Thompson\label{tab:data}} 
\tablecomments{Compilation of distance-vs-time observations for Usain
  Bolt and Richard Thompson in the 100 meter dash in Beijing 2008, obtained
  from screen shot prints of the race. }
\tablecolumns{7}
\tablehead{Uncalibrated  & \multicolumn{2}{c}{{\bf Usain Bolt}} &
  \multicolumn{2}{c}{{\bf Richard Thompson}} & 
  &  \\ 
elapsed time & Ticks & Distance & Ticks & Distance & Uncertainty & Data set \\
(s) & (\#) & (m)& (\#) & (m) & (m)& 
}
\startdata
0.0\tablenotemark{*}  &     -7.0   & 0.0 &    -7.0    & 0.0 &
0.0 &
None \\
(0.01 &    -7.0   & 0.0 &    -7.0    & 0.0 & 0.0 &None)\tablenotemark{$\dagger$}\\
1.1  &     -2.0   & 5.0 &    -2.1    & 4.9 &  0.5 &       NRK\\
3.0  &     15.5   & 22.5 &    15.6    & 22.6 &  0.5 &       NRK\\
4.0  &     27.0   & 34.0 &    27.0    & 34.0 &  0.4&       NRK\\
4.5  &     34.3   & 41.3 &    34.1    & 41.1 &  0.5&       NRK\\
5.4  &     45.1   & 52.1 &    44.3    & 51.3 &  0.5&       NBC\\
5.8  &     48.9   & 55.9 &    48.3    & 55.3 &  0.5&       BBC\\
6.2  &     54.5   & 61.5 &    53.8    & 60.8 &  0.5&       NBC\\
6.5  &     57.8   & 64.8 &    56.9    & 63.9 &  0.4&       BBC\\
6.9  &     62.6   & 69.6 &    61.5    & 68.5 &  0.2&       NBC\\
7.3  &     66.3   & 73.3 &    65.1    & 72.1 &  0.2&       NBC\\
7.7  &     71.5   & 78.5 &    70.1    & 77.1 &  0.2&       NBC\\
8.0  &     74.7   & 81.7 &    72.9    & 79.9 &  0.2&       NBC\\
8.3  &     78.6   & 85.6 &    76.8    & 83.8 &  0.2&       NBC\\
8.6  &     82.2   & 89.2 &    80.5    & 87.5 &  0.2&       NBC\\
8.8  &    84.3   & 91.3 &    82.4    & 89.4 &  0.2&       NBC      \\
9.4  &     91.6   & 98.6 &    89.4    & 96.4 &  0.2&       NBC\\
9.69\tablenotemark{*}  &   93.0   & 100. &    \nodata    & \nodata & 0.0&        NRK\\
9.89\tablenotemark{*}  &   \nodata   & \nodata &    93.0    & 100. & 0.0&        NRK\\
(13   &    105     & 112. &  105.      & 112 &       5.0 &NRK)\tablenotemark{$\dagger$}
\enddata
\tablenotetext{*}{The first point is taken from the known starting
  position, and the last is taken from a high-resolution picture of
  the finishing line. The times for these are not read from the screen
clock, but are adopted from official sources. Zero uncertainties are
assigned to these points.} 
\tablenotetext{$\dagger$}{These two points are not real observations, but
  auxiliary points to ensure sensible boundary conditions for the
  smooth spline; the first ensures zero starting velocity, and the
  last gives a smooth acceleration at the finishing line.}
\end{deluxetable*}

\begin{figure}
\mbox{\epsfig{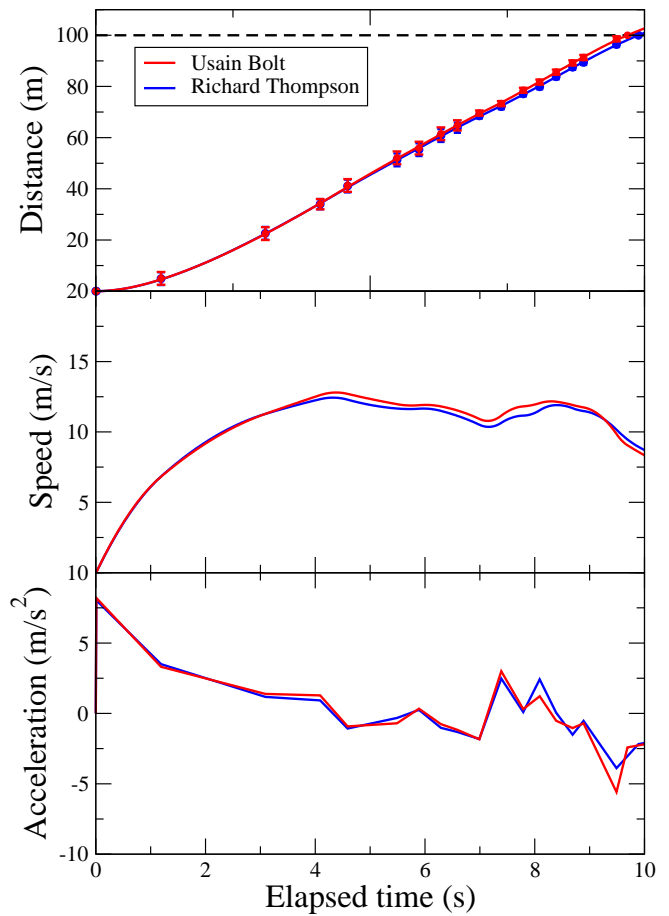}}
\caption{Estimated position (top), speed (middle) and acceleration
  (bottom) for Bolt (red curves) and Thompson (blue curves) as a
  function of time. Actual distance measurements are indicated in the
  top panel with $5\sigma$ error bars.}
\label{fig:evolution}
\end{figure}

\section{Data and observations}
\label{sec:data}

\subsection{Data sets}

The data used for this analysis consist of three clips filmed by three
cameras located along the finishing line at slightly different
positions. Specifically, the clips were obtained from NRK, NBC, and
BBC.

Unfortunately, the NRK and BBC clips were filmed with cameras
positioned fairly close to the track, and the rail of the moving
camera therefore disappears outside the field-of-view after about 6
seconds. This is not the case for the NBC clip, which was filmed from
further away. Even though the quality of this version is rather poor,
it is possible to count the number of ticks to the end. 

\begin{figure*}
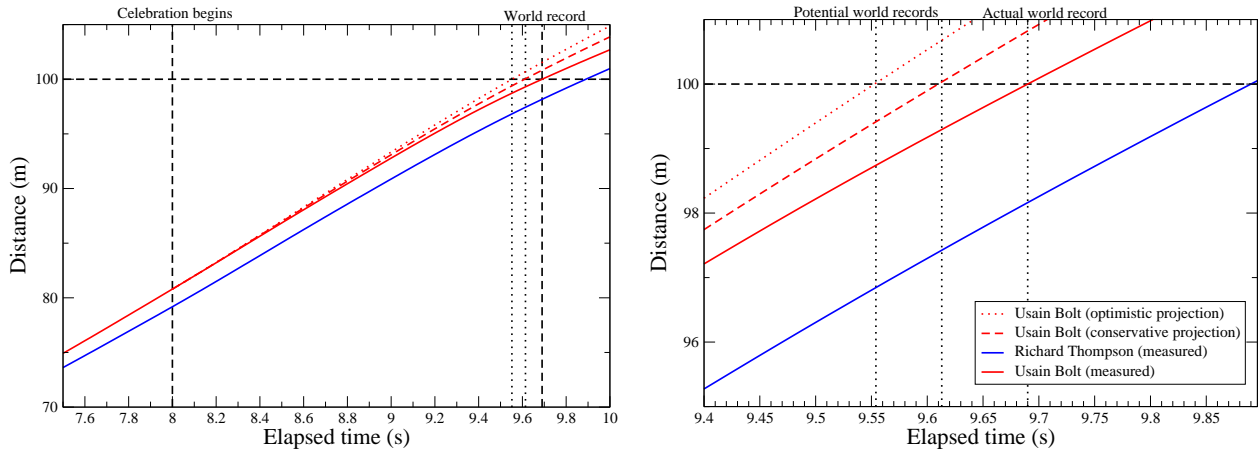

\mbox{\subfigure{\label{fig:end_run_a}\epsfig{figure=projected_end_run_corrected.eps,width=0.45\textwidth,clip=}}
      \quad
      \subfigure{\label{fig:end_run_b}\epsfig{figure=projected_end_run_zoom_corrected.eps,width=0.45\textwidth,clip=}}}
    \caption{Comparison of real and projected distance profiles at the
      end of the race. The point where the profiles cross the
      horizontal 100 meter line is the new world record for a given
      scenario. The right panel is only a zoomed version of the left
      panel.}
\label{fig:end_run}
\end{figure*}

Using these data sets, we measured the position of Usain Bolt and
Richard Thompson at 16 different times in units of ticks. These are
all listed in Table \ref{tab:data}.

\subsection{Calibration of measurements}

There are three issues that must be addressed before the tick counts
listed in Table \ref{tab:data} can be translated into proper distance
measurements. First, the camera rail is not visible entirely to the
starting line, as the very first part is obscured by a camera man. The
tick counts in Table \ref{tab:data} are therefore counted relative to
the first visible tick. Fortunately, it is not very problematic to
extrapolate into the obscured region by using the distance between the
visible ticks, and knowing that the distance between the starting
lines for the 100 meter dash and 110 meter hurdles is precisely 10
meters. We estimate the number of obscured ticks to be $7 \pm 1$.

Second, the precision of the screen clock is only a tenth of a second,
and the clock also appears to truncate the time, not round off. We
therefore add 0.05 seconds to each time measurement, and define our
uncertainty in time to uniform between -0.05 and 0.05 seconds.

Finally, the screen clock is not calibrated perfectly with the stadium
clock. (See Figure 1 for an example frame.) A little more than half of
all frames appear to be synchronized, while in the rest the screen
clock is lagging behind by 0.1 seconds. We assume that the stadium
clock is the correct one, and re-calibrate the screen clock by adding
an additional 0.04 seconds to each time measurement.

With these assumptions, it is straightforward to calibrate both the
clock and distance measurements, and this is done in the corresponding
columns in Table \ref{tab:data}.

\section{Estimation of motion profiles}
\label{sec:analysis}

With calibrated distance information ready at hand, it is
straightforward to make the desired predictions. First, we compute a
smooth spline \citep{green:1994}, $s(t)$, through each of the two
runners' measured positions. A nice bonus of using splines is that we
automatically obtain the second derivatives of $s$ (ie., acceleration)
at each time step, and also the first derivatives (ie., speed),
\begin{equation}
v(t) = \frac{ds}{dt}; \quad a(t) = \frac{dv}{dt} = \frac{d^2 s}{dt^2}.
\label{eq:v_and_a}
\end{equation}

To obtain a well-behaved spline, we impose three constraints. First,
we add two auxiliary data points at $t=0.01$ and $t=13.0$
seconds. These are not measurements, but included only in order to
guarantee sensible boundary conditions at each end: The first one
implies that the starting velocity is zero, while the last one leads
to a smooth acceleration at the finishing line. Thirdly, we adopt a
smooth spline stiffness parameter of $\alpha=0.5$ \citep{green:1994}
to minimize unphysical fluctuations. The results are fairly
insensitive to the specific value of this parameter.

The resulting functions are plotted in Figure
\ref{fig:evolution}. Some interesting points to notice are the
following:
\begin{itemize}
\item Bolt and Thompson are virtually neck by neck up to four seconds,
  corresponding to a distance of 35 meters. 
\item Bolt's Olympic gold medal is essentially won between 4 and 8
  seconds. 
\item At 8 seconds Bolt decelerates noticeably, and Thompson equalizes
  and surpasses Bolt's speed. Note however that Thompson is also not
  able to maintain his speed to the very end, but runs out of power
  after about 8.5 seconds. Still, his acceleration is consistently
  higher than Bolt's after 8 seconds.
\end{itemize}

\section{World record projections}
\label{sec:projections}

We are now in the position to quantitatively answer the original
question: How fast would Bolt really have run, if he hadn't celebrated
the last 2 seconds? To make this projection, we consider the following
two scenarios:
\begin{enumerate}
\item Bolt matches Thompson's acceleration profile after 8 seconds.
\item Bolt maintains a 0.5 m/s$^2$ higher acceleration than Thompson
  after 8 seconds.
\end{enumerate}
The justification of scenario 1 is obvious, as Bolt outran Thompson
between 4 and 8 seconds. The justification of scenario 2 is more
speculative, as it is difficult to quantify exactly \emph{how much}
stronger Bolt was. Still, looking at the acceleration profiles in
Figure \ref{fig:evolution}, and noting that Bolt traditionally was
considered a 200 meter specialist, a value of 0.5 m/s$^2$ seems fairly
realistic.

Then, for each scenario we compute a new trajectory for Bolt by
choosing initial conditions, $s_0 = s(8 \,\textrm{sec})$ and $v_0 =
v(8\,\textrm{sec})$, and an acceleration profile as described
above. The computation of these trajectories are performed by simply
integrating Equation \ref{eq:v_and_a} with respect to time,
\begin{align}
\hat{s}(t) &= s_0 + \int_{t_0}^{t} \hat{v}(t) dt \\ 
\hat{v}(t) &= v_0 + \int_{t_0}^{t} \hat{a}(t) dt \\ 
\hat{a}(t) &= a_{\textrm{Thompson}}(t).
\end{align}

In Figure \ref{fig:end_run} we compare the projected trajectories,
$\hat{s}(t)$ (dashed red line shows scenario 1, dotted red line shows
scenario 2), with the actual trajectory, $s(t)$ (solid red line). For
comparison, Thompson's trajectory is indicated by a solid blue
line. 

The projected new world record is the time for which $\hat{s}(t)$
equals 100 meter. Including 95\% statistical errors estimated by Monte
Carlo simulations as described in Section \ref{sec:method}, we find
that the new world record would be $9.61\pm0.04$ seconds in scenario
1, and $9.55\pm0.04$ in scenario 2.

\section{Conclusions}
\label{sec:conclusions}

Glen Mills, Usain Bolt's coach, suggested that the world record could
have been 9.52 seconds if Bolt had not danced along the track in
Beijing for the last 20 meters. According to our calculations, that
seems like an good, but perhaps slightly optimistic, estimate:
Depending on assumptions about Bolt's acceleration at the end of the
race, we find that his time would have been somewhere between 9.55 and
9.61 seconds, with a 95\% statistical error of $\pm0.04$
seconds. Clearly, the uncertainties due to the assumptions about the
acceleration are comparable to or larger than the statistical
uncertainties. Therefore, 9.52 seconds does by no means seem to be out
of reach.

\begin{figure}
\mbox{\epsfig{file=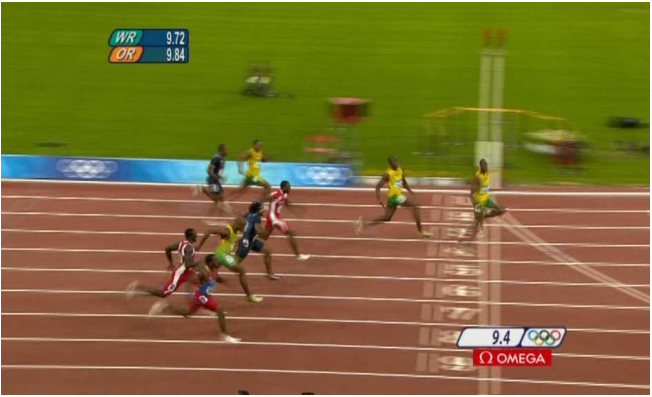,width=\linewidth,clip=}}
\caption{Photo montage showing Bolt's position relative to his
  competitors for real (left Bolt) and projected (right Bolt) world
  records.}
\label{fig:manipulated}
\end{figure}

In Figure \ref{fig:manipulated} we show an illustration of how such a
record would compare to the actual world record of 9.69 seconds,
relative to the rest of the field: The left version of Bolt shows his
actual position at $\sim9.5$ seconds, while the right version
indicates his position in the new scenarios.

Of course, there are potential several systematics errors involved in
these calculations. For instance, it is impossible to know for sure
whether Usain might have been tired at the end, which of course would
increase the world record beyond our estimates. On the other hand,
judging from his facial expressions as he crossed the finishing line,
this doesn't immediately strike us as a very plausible hypothesis.

Another issue to consider is the wind. It is generally agreed that a
tail wind speed of 1 m/s improves a 100 meter time by 0.05 seconds
\citep{mureika:2000}. Further, for IAAF (International Association of
Athletics Federations) to acknowledge a given run as a record attempt,
the wind speed must be less than +2 m/s. When Bolt
ran in Beijing, there was no measurable wind speed at all, and one can
therefore safely assume that the world record could have been further
decreased, perhaps by as much as 0.1 seconds, under more favorable
wind conditions.

A corollary of this study is that a new world record of less than 9.5
seconds is within reach for Usain Bolt in the near future.

\begin{acknowledgements}
  First and foremost, we would like to thank Christian Nitschke Smith
  at NRK Sporten for providing very useful high-resolution footage of
  the Beijing 100 dash run, and also BBC and NBC for making their
  videos available on their web pages. Second, we thank the pizza guy
  from Peppe's who provided us with a very good half-n-half ``Thai
  Chicken'' and ``Heavy Heaven'' pizza on a late Friday night. This
  article has been submitted to the American Journal of
  Physics. After it is published, it will be found at
  http://scitation.aip.org/ajp.
\end{acknowledgements}

\end{document}